# Electron interferometry in quantum Hall regime: Aharonov-Bohm effect of interacting electrons

Ping V. Lin[1], F. E. Camino[2], and V. J. Goldman[1]

[1] *Department of Physics, Stony Brook University, Stony Brook, NY 11794-3800, USA*
[2] *Center for Functional Nanomaterials, Brookhaven National Laboratory, Upton, NY 11973, USA*

An apparent $h/fe$ Aharonov-Bohm flux period, where $f$ is an integer, has been reported in coherent quantum Hall devices. Such sub-period is not expected for non-interacting electrons and thus is thought to result from interelectron Coulomb interaction. Here we report experiments in a Fabry-Perot interferometer comprised of two wide constrictions enclosing an electron island. By carefully tuning the constriction front gates, we find a regime where interference oscillations with period $h/2e$ persist *throughout* the transition between the integer quantum Hall plateaus 2 and 3, including half-filling. In a large quantum Hall sample, a transition between integer plateaus occurs near half-filling, where the bulk of the sample becomes delocalized and thus dissipative bulk current flows between the counterpropagating edges ("backscattering"). In a quantum Hall constriction, where conductance is due to electron tunneling, a transition between forward- and back-scattering is expected near the half-filling. In our experiment, neither period nor amplitude of the oscillations show a discontinuity at half-filling, indicating that only one interference path exists throughout the transition. We also present experiments and an analysis of the front-gate dependence of the phase of the oscillations. The results point to a single physical mechanism of the observed conductance oscillations: Aharonov-Bohm interference of interacting electrons in quantum Hall regime.

The Aharonov-Bohm effect demonstrates the primacy of the potentials, rather than fields in quantum mechanics.[1-3] Specifically, for a dilute beam of non-interacting electrons propagating in a magnetic field $\mathbf{B}$, the vector potential $\mathbf{A}$ attaches a phase factor $\exp\{-i\frac{e}{\hbar}\int_O^{\mathbf{r}}\mathbf{A}(\mathbf{r'})\cdot d\mathbf{r'}\}$ to the electron wave function at position $\mathbf{r}$. For closed electron orbits, the phase factor is periodic in flux $\Phi$ through the area $S$ enclosed by the interference path: $\Phi = \oint \mathbf{A}(\mathbf{r})\cdot d\mathbf{r} = \int_S \mathbf{B}\cdot d\mathbf{S}$ by the virtue of the Stokes' theorem; the $2\pi$ period of the phase corresponds to the $\Phi_0 \equiv 2\pi\hbar/e = h/e$ flux period.

Electron interaction usually does not affect the $h/e$ Aharonov-Bohm flux period observed in conductance of normal metal and semiconductor rings with two leads. The situation is more complex in quantum Hall devices. An apparent $h/fe$ Aharonov-Bohm flux period, where $f$ is the integer quantum Hall effect (QHE) filling in the constrictions, has been reported in quantum antidot[4,5] and Fabry-Perot interferometer[6-8] devices. In quantum antidots, the closed Aharonov-Bohm path follows an equipotential around the lithographically-defined potential hill in the two-dimensional (2D) electron plane. In interferometer devices, the interference path follows an equipotential at device's edges, and is closed by two tunneling links.

The experiments are done in a uniform magnetic field, so that a well-defined interference path enclosing an area is needed to translate the field into flux. This Aharonov-Bohm sub-period is accompanied by an $e$ charge period as a function of a gate voltage, and is not affected by the 2D bulk filling outside the device. In quantum antidots, previously reported $h/2e$ period[9,10] was tentatively attributed to spin-splitting of a Landau level. However, subsequent work has concluded that no model of non-interacting electrons can consistently explain this sub-period.[11,12] On the other hand, it seems apparent that the strong interelectron Coulomb interaction, present in nearly all QHE samples, can naturally cause the observed Aharonov-Bohm and charge periods by substantially mixing the Landau level electron occupation.[5]

An isolated metallic island weakly coupled by tunneling to two electrodes displays quasi-periodic conductance oscillations observed as a function of gate voltage. In such Coulomb islands,[13,14] the net island charge $Q = -e(N - N_{eq})$ increments in steps of one electron due to the Coulomb blockade which opens a gap of $Q^2 / 2C$ in the island energy spectrum. The island has total capacitance $C$ to the gate and the electrodes. Here $N$ is the number of electrons in the nearly isolated island, an integer, and the equilibrium expectation value $N_{eq} = N_{ion} + N_{gate}$ is the sum of two terms: the number of electrons neutralizing the positively-charged background of the fixed ions in the crystal lattice $N_{ion}$ and the continuously-varying polarization charge $N_{gate} = -\alpha V_{gate}$ induced by a gate voltage $V_{gate}$. Under conditions of low temperature and excitation (bias voltage between the two electrodes), the net island charge oscillates between $Q = -\frac{1}{2}e$ and $Q = \frac{1}{2}e$, conductance peaks occurring at gate voltages when $N_{eq}$ is an integer and $Q$ is zero, so the Coulomb gap vanishes.

Phenomenological Coulomb blockade models were proposed to evaluate the effects of on-site interaction in quantum antidot[11,12] and Fabry-Perot geometry.[15-19] Specifically, it has been proposed that two distinct mechanisms producing conductance oscillations exist: one being Aharonov-Bohm interference of back-scattered electrons, another is caused by forward-scattering via a "compressible island" subject to Coulomb blockade (see Fig. 1 in Ref. 17). The third possibility, the backscattering via a "compressible island" (shown in Fig. 1 in Ref. 20 and as "type ii" in Ref. 17), does not conserve angular momentum in the integer QHE regime, and thus is expected to be much weaker. Experiments aimed at distinguishing the distinct Coulomb blockade and the Aharonov-Bohm mechanisms have been reported.[21-23]

However: (i) a "compressible island" has no well-defined area, so that while Coulomb blockade is possible, it does not necessarily lead to $B$-periodic oscillations as a function of a uniform applied magnetic field. (ii) A "compressible island", if formed, would vary in size and shape from a point at the island center to a ring of maximal radius, when filling is changed in a QHE plateau transition; thus the forward tunneling distance, and tunneling conductance, would vary enormously. (iii) Further, in semiconductor heterostructures with ~200 nm 2D depletion length, the confining potential has considerable radial gradient which results in a discrete island energy spectrum; no strictly compressible island is possible in the limit of low temperature.

The state of affairs is further obscured by the fact that the single-electron tunneling dynamics is similar for the discrete electron spectra resulting from Coulomb blockade and quantum confinement. In particular, the Schrödinger equation can be solved for an electron constrained to move on a circular ring of radius $R$ enclosing flux $\Phi$. The energy is periodic in $\Phi$, $E = \frac{e^2}{8\pi^2 mR^2}(\Phi - n\Phi_0)^2$, where $n$ is an integer. The lowest energy radii correspond to enclosed flux of an integer multiple of $\Phi_0$. If the orbit radius is fixed and the applied magnetic field is varied, this Aharonov-Bohm periodic energy dependence, consisting of a set of intersecting parabolas, with the ground state switching at half-integer values of $\Phi/\Phi_0$, is similar to the Coulomb blockade energy $E = \frac{e^2}{2C}(N - N_{eq})^2$. Thus, the characteristic tunneling conductance "Coulomb blockade diamonds" seen in the source-drain bias versus gate voltage plots[11,23] are also expected for any size-quantized electron system with a discrete energy spectrum, including an Aharonov-Bohm ring.

Here we report experiments on a Fabry-Perot electron interferometer in the regime of transition between $f = 2$ and 3 QHE plateaus. By fine-tuning the two constrictions, we have obtained a continuous sequence of the Aharonov-Bohm oscillations persisting *throughout* the transition, including Landau level filling $\nu = 2.5$. The half-filling $\nu = f + \frac{1}{2}$ separates the high-$B$ side of the $f + 1$ plateau and the low-$B$ side of the $f$ plateau.[24,25] The two situations have been interpreted as corresponding to back- and forward-scattering regimes, respectively.[17,21] We observe experimental flux period $h/2e$ all through the plateau transition, although a period of $h/3e$ is expected for the $f = 3$ plateau. We also present experiments and an analysis of the gate dependence of the phase of the oscillations that shows that the slope of the constant-phase stripes depends on details of the confining potential and device geometry. We conclude that all reported experimental results can be understood without invoking tunneling via a "compressible island". The observed continuous sequence of sub-$h/e$ period oscillations argues strongly for a single physical mechanism: the Aharonov-Bohm interference of interacting electrons in QHE regime.

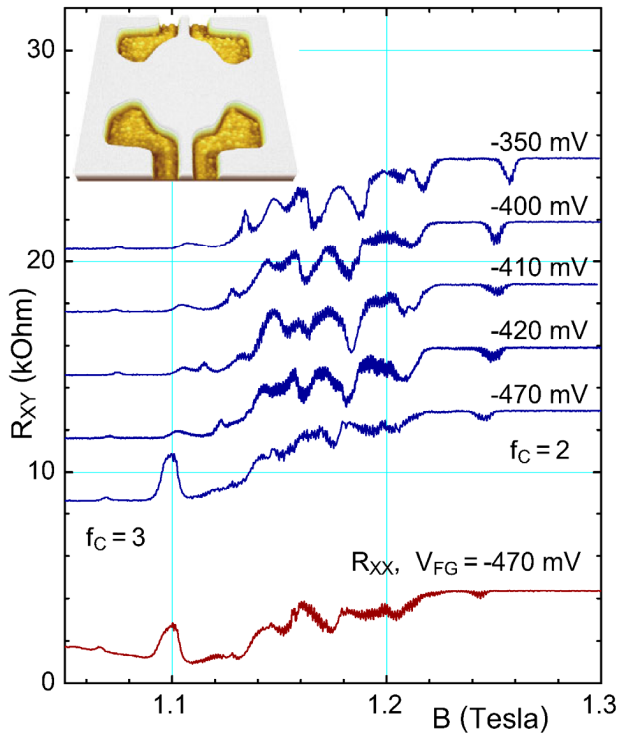

FIG. 1. The Hall ($R_{XY}$) and longitudinal ($R_{XX}$, lowest trace) resistance of the interferometer device between constriction $f = 3$ and 2 QHE plateaus. The successive $R_{XY}$ traces are shifted by 3 k$\Omega$ and are labeled by bias of one of the front gates, the other three voltages are constant. Inset shows a $4 \times 4$ $\mu$m Atomic Force micrograph of the central region of the device.

The Fabry-Perot device, shown in the inset in Fig. 1, was described previously.[26,27] The etch trenches define two 1.2 $\mu$m-wide constrictions, which separate an approximately circular electron island from the 2D bulk. Tunneling occurs in the two constrictions, thus forming a Fabry-Perot interferometer. The depletion potential of the trenches determines the electron density profile, see Fig. 1 in Ref. 26. Four Au/Ti front gates are deposited in the etch trenches. Front gates are used to fine-tune the constrictions for symmetry of the tunneling and to vary the overall device electron density, but the shape of the electron confinement potential is dominated by the etch trench depletion. The 2D density $\sim 1 \times 10^{11}$ cm$^{-2}$ is achieved by illumination at 4.2K, there are $\sim 3,000$ electrons in the island. Four-terminal longitudinal $R_{XX}$ and Hall $R_{XY}$ resistances (see inset in Fig. 2) were measured with 200 or 400 pA, 5.4 Hz AC current excitation. All data reported here were taken at the bath temperature of 10 mK.

Figure 1 shows several $R_{XY}$ traces, each with slightly different front-gate voltage on one side of one constriction. The $f = 2$ and 3 constriction plateaus are connected by a QHE transition region, where the $h/2e$ Aharonov-Bohm oscillations are superimposed on a varying background. Similar oscillations are also seen in $R_{XX}$. In general, unless the two constrictions are fine-tuned, the $B$-regions with oscillations are interrupted, so that the plateau transition does not contain a continuous oscillation sequence. In a large 2D sample, a transition between two plateaus displays a smooth, monotonic $R_{XY}$.[28] The aperiodic peaks or dips in Fig. 1, spaced by ~0.03 T, are attributed to disorder-assisted tunneling outside the constrictions; similar ubiquitous mesoscopic fluctuations are also seen in the same device at lower magnetic fields[27] and in quantum antidots. That the aperiodic peaks originate outside of the island is evidenced by their different response to front gates of the left and right constrictions.

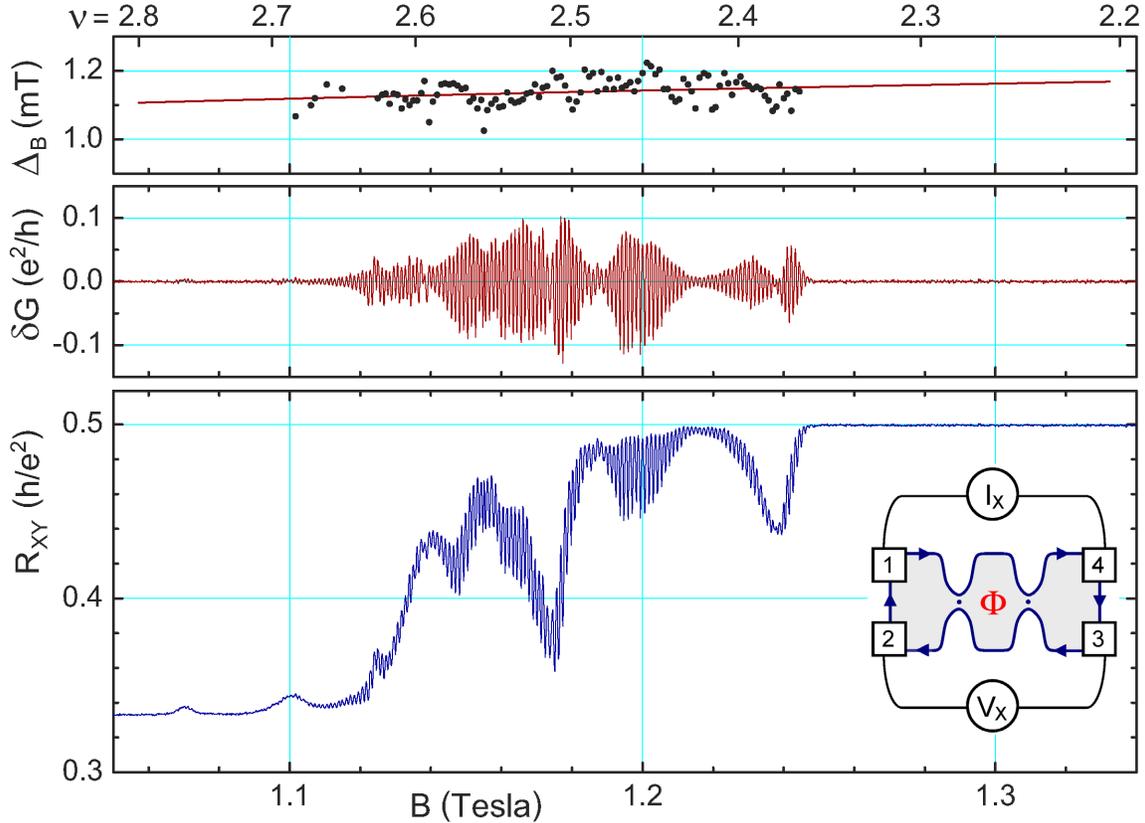

FIG. 2. Hall resistance of the interferometer for the 2↔3 QHE plateau transition, the Landau level filling is given at the top. The middle panel shows the oscillatory conductance, and the upper panel the oscillation period; the red line is the linear fit: $\Delta_B = 0.104(2.5 - \nu) + 1.14$ mT. The oscillations persist uninterrupted *throughout* the transition region, including the half-filling. The inset shows four-terminal measurement configuration for $R_{XX} = V_X / I_X$; for $R_{XY}$ current is passed 1-3, voltage is measured on contacts 2-4.

By fine-tuning the constriction gate voltage, we managed to obtain a continuous, uninterrupted oscillation sequence. Figure 2 shows a high-resolution $R_{XY}$ trace measured with 200 pA excitation, and the oscillatory conductance $\delta G$ obtained by subtracting a smooth background. Although 200 pA produces only 2 μV constriction Hall voltage at $\nu \approx 2.5$, we still observe non-Ohmic behavior, namely, the oscillatory conductance amplitude still increases upon lowering of the excitation at 10 mK temperature. This is evidence that the extensive cold filtering employed lowers the electromagnetic background "noise" to $\leq 2$ μV at the sample's contacts.

Note that the conductance oscillations can be seen, without interruption, throughout the transition from the $R_{XY} = \frac{1}{3}\frac{h}{e^2}$ to the $\frac{1}{2}\frac{h}{e^2}$ QHE plateaus. The top panel of Fig. 2 shows the magnetic field period of the oscillations . In these data $\Delta_B = 1.14$ mT is closely one-half of the 2.3 mT $f = 1$ period. Thus we interpret the oscillatory data in Fig. 2 as displaying two oscillations per $h/e$, the fundamental flux period, in agreement with earlier results.[6-8,20-22,26] The same $h/2e$ flux period persists in the whole $2 \leftrightarrow 3$ QHE transition region. The weak, systematic variation of $\Delta_B$ as $B$ is increased is caused by the gradual, secular inward shift of the island-circling edge channel (interference path area shrinks), so as to maintain a constant $\nu = hn/eB$ in the local edge-channel electron density $n$ when $B$ is changing. The sign of the $d\Delta_B/dB$ slope is consistent with both: forward- and back-scattering at the saddle point in the constrictions. The oscillation amplitude is maximal near half-filing, and falls off towards the quantized plateaus, similar to that reported in a Mach-Zehnder interferometer.[29]

We discuss these data in terms of a specific edge-channel model below. Here we note that the oscillatory behavior in Fig. 2 is dramatically different from resistance peaks and dips in quantum antidots. In quantum antidots, resonant tunneling peaks are seen on the low-$\nu$ side of a QHE plateau, and dips on the high-$\nu$ side of the *same* plateau, both having equal flux period $h/fe$.[5] In particular for the $2 \leq \nu \leq 3$ transition, there are *two* dips per $h/e$ below the $f = 2$ plateau, and *three* peaks per $h/e$ above the $f = 3$ plateau, separated by a smooth region near half-filling. Such behavior is consistent with two distinct tunneling regimes of back- and forward-scattering in the antidot geometry. The continuous oscillation sequence with a constant flux period is consistent with only back-scattering occurring in Fabry-Perot interferometers.

Figure 3 shows conductance oscillations at $\nu \approx 2.36$ with the front-gate voltage $V_{FG}$ as a parameter. Here, all four $V_{FG1-4}$ are stepped by a common bias of 0.10 mV, and the average $V_{FG} = \frac{1}{4}\sum_j V_{FGj}$. The 2D electron density is greater in this cooldown than in Fig. 2, so that equal $\nu$ occurs at a higher $B$. The fundamental flux period $h/e$ contains two conductance oscillations, $S\Delta_B = h/2e$. Stepping $V_{FG}$ more negative reduces the overall island electron density and thus shifts the region of oscillations to lower $B$, see Fig. 3(b). The flux period is constant, but $V_{FG}$ changes the interference path area $S$, see Fig. 3(b), thus changing the $B$-field period $\Delta_B$.[6,8] The constant phase of oscillations form stripes spaced vertically by $\Delta_{V_{FG}} = 0.7 \pm 0.1$ mV. Interpreting $\Delta_{V_{FG}}$ as matching the change of the number of electrons within $S$ by one gives $S(dn/dV_{FG})\Delta_{V_{FG}} = 1.0$, using the experimental $dn/dV_{FG} = 7.9 \times 10^{14}$ m$^{-2}$V$^{-1}$, obtained from the low-$B$ magnetotransport,[27] and $S = h/2e\Delta_B = 1.83 \times 10^{-12}$ m$^2$, obtained from the Aharonov-Bohm period in Fig. 3. This satisfactory agreement supports validity of our interpretation.

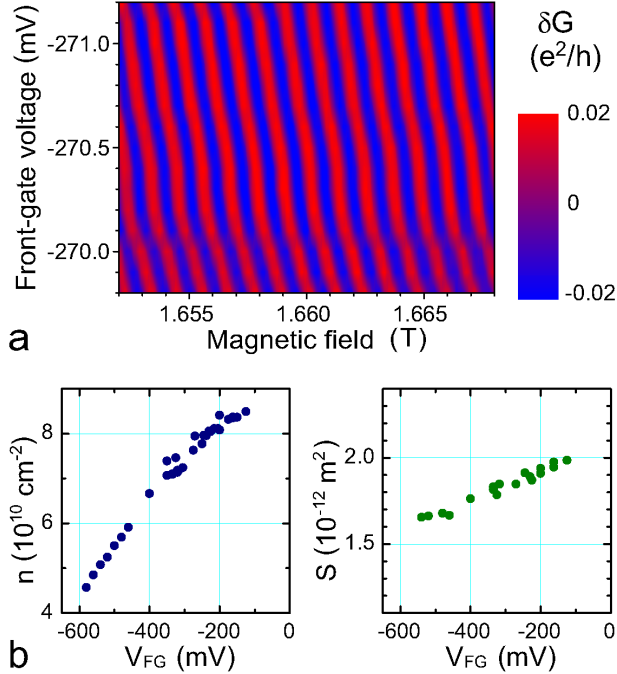

FIG. 3. (a) 3D color plot of the Aharonov-Bohm oscillations on the $f = 2$ plateau ($\nu \approx 2.36$). A negative $V_{FG}$ is stepped by 0.10 mV. The slope of the constant oscillation phase stripes is positive, consistent with Aharonov-Bohm effect in a QHE interferometer, as discussed in the text. (b) Constriction electron density and interference path area dependence on front-gate voltage. Constriction $n$ and interference $S$ are determined from the $B$-field position and period of the $f = 2$ oscillations ($\nu \approx 2.5$) in several cooldowns.

In general, the sign of the constant phase slope depends on details of heterostructure material and device geometry and fabrication. For one electron the Aharonov-Bohm phase $\gamma = -e\Phi/\hbar$. In a uniform $B$, the flux through the interference path $\Phi = BS$, and the differential

$$d\gamma/2\pi = -(e/h)(SdB + BdS).\qquad(1)$$

In the QHE of non-interacting electrons, in the symmetric gauge, each orbital in each Landau level is quantized so as to enclose an integer multiple of $\Phi_0 = h/e$,[24,25] this also minimizes the

$$\frac{e^2}{8\pi^2 mR^2}(\Phi - n\Phi_0)^2$$ per electron Aharonov-Bohm energy. Thus, the QHE ground-state maximum density electron droplet (a completely filled Landau level) is constructed by filling the Aharonov-Bohm orbitals from the center of the island (the minimum of the confining potential) outwards. Likewise, a partially-filled Landau level contains an integer number of electrons within an Aharonov-Bohm path. Therefore, even the orbitals of non-interacting electrons in QHE regime are quantized to enclose an integer number of electrons in each Landau level, and invoking Coulomb blockade in ill-defined areas to ensure an integer number of electrons is *redundant* in this open geometry.

Between the QHE plateaus, at filling $f < \nu < f + 1$, when $f$ Landau levels are completely filled, lowering the uniform magnetic field and thus $SdB$ by $h/e$ "excites" $f$ electrons. Thus $SdB = -(h/ef)dN_e$, where $N_e$ is thermal average number of the excited electrons (electrons in the $f + 1^{st}$ Landau level) enclosed by the path. One may argue that the excitation of an electron into the partially-filled Landau level is likely to modulate the conductance via the interference path closed by tunneling, and thus result in conductance oscillations. This explains why there may be $f$ conductance oscillations within the fundamental $h/e$ period (or tunneling peaks in quantum antidots), but does not explain why the $f$ oscillations are equally spaced in $B$ or have equal amplitude.[5] Indeed, for *non-interacting* electrons the positions of oscillations in $B$ depend on the detail of the confining potential, each oscillation originating in a different filled Landau level. For *interacting* electrons, the many-electron ground states involve occupation of higher Landau levels, "Landau level mixing". But the basis orbitals, and thus the interference paths, are still quantized by the Aharonov-Bohm flux condition. When electron-electron interaction is strong, occupation of neighboring Landau levels is similar, and the "excited" electrons do not originate in any specific Landau level. Thus, excitation of an electron by reduction of $B$ would result in approximately equivalent oscillations. This provides a qualitative model explaining the experimental observations as resulting from effects of electron Coulomb interaction on Aharonov-Bohm effect in QHE regime.[5] However, this qualitative model has proven difficult to implement in a formal theory.

We now turn to consideration of the effect of front gates. For interacting electrons, minimization of the ground state energy requires *local* charge neutrality for 2D density $n$.[24,25] Thus $\partial n / \partial B = 0$ and $\partial n_e / \partial B = -ef / h$. Using $B$ and a gate voltage $V$ as two independent variables, the differentials $dN_e = (\partial N_e / \partial B)dB + (\partial N_e / \partial V)dV$ and $dS = (\partial S / \partial B)dB + (\partial S / \partial V)dV$. Here, the total number of electrons within the interference path is $N = nS$, $\partial N_e / \partial B = -feS / h + (1 - f / \nu)n(\partial S / \partial B)$ and $\partial N_e / \partial V = S(\partial n / \partial V) + (1 - f / \nu)n(\partial S / \partial V)$ because the fraction of the "excited" electrons is $1 - f / \nu$, $n_e = (1 - f / \nu)n$. A gate changes the occupation only of the partially-filled Landau level: $\partial n / \partial V = \partial n_e / \partial V$ in a fixed $B$. Combining the terms and defining $\beta = 2f / \nu - 1$, we obtain

$$\frac{f\,d\gamma}{2\pi} = -\left[\frac{feS}{h} + \beta n\frac{\partial S}{\partial B}\right]dB + \left[S\frac{\partial n}{\partial V} - \beta n\frac{\partial S}{\partial V}\right]dV .\qquad(2)$$

For etch trench depletion a good approximation may be the hard confinement: $\partial S / \partial B = 0$, $\partial S / \partial V = 0$. Then the periods (exciting one electron, $f\Delta_\gamma = 2\pi$) are $S\Delta_B = h / ef$ and $S\Delta_V = (\partial n / \partial V)^{-1}$. Note that the $B = const$, $V = const$ partial derivatives in Eq. 2 are not equal to the experimental slopes in Fig. 3(b), which correspond to $\nu \approx const$.

The $d\gamma = 0$ stripe slope depends on the signs of the $dB$ and $dV$ multipliers in Eq. (2). For electrons, the chief $dB$ term is always positive. The net sign of the $dV$ term depends on the two contributions. Positive gate voltage attracts electrons: $\partial n / \partial V$ is always positive; $\partial S / \partial V$ is negative for anticonfining (quantum antidots[4,5]) and positive for confining potential (Fabry-Perot devices). Mach-Zehnder devices[29-31] have one edge with confining and one with anticonfining potential, the net term depends on device details. In most experiments $\beta \sim 1$. Thus, the hard confinement model predicts a small positive $\gamma = const$, $dV / dB$ slope for Fabry-Perot interferometers and quantum antidots. For devices with soft confinement and/or modulation gates the $dV / dB$ slope can be large in magnitude (weak net gate coupling), and its sign depends on device details. A small modulation gate may have $Sdn$ and $ndS$ effects different than large gates.

A Coulomb blockade model of Ref. 17 was used in Ref. 22. It predicts constant oscillation phase when the charging energy is constant, $dN_e = 0$. However, the island area is assumed fixed for one device, while not so for the larger, very constricted ($n_C / n_B = 0.4$) device. We see that, in general, there are more terms contributing: gate voltage changes flux, too, by affecting the area ($\nu = const$ means $n / B = const$, not $N / B = const$).

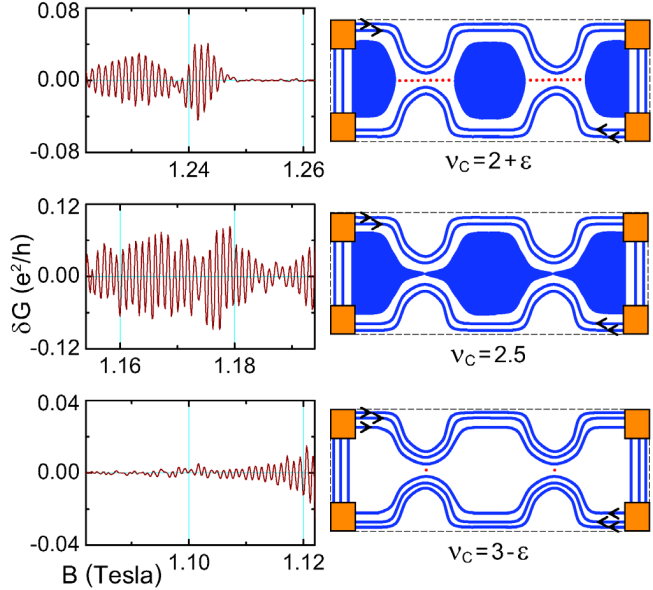

FIG. 4. Right: illustration of edge-channel structure in a Fabry-Perot interferometer for the 2↔3 QHE plateau transition. The arrowed lines show edge channels connecting the Ohmic contacts (squares). Red dots in the constrictions show tunneling. The three regimes illustrated correspond to back- ($\nu_C = 3 - \varepsilon$) and forward- ($\nu_C = 2 + \varepsilon$) scattering, and to half-filling. Left: the corresponding experimental oscillatory conductance for the 2↔3 QHE transition, see Fig. 2.

When the interference oscillations are observed, the counterpropagating edge channels must pass near the saddle points in the constrictions where tunneling occurs. Thus the filling of the relevant edge channels is determined by the saddle point filling $\nu_C$. The filling outside the constrictions and at the island center $\nu_I$ is greater than $\nu_C$, the exact profile of the depletion is determined by the heterostructure material, device geometry and fabrication, and also by the gate voltage. In this device $n_I / n_C \approx 1.07$, thus the entire interferometer, including the island center and the constrictions, is on the same integer QHE plateau for $f \leq 8$.[26,27]

Figure 4 shows an illustration of edge-channel configurations used to analyze the Fabry-Perot geometry in Ref. 17. In this model, lines represent compressible edge channels, where local filling varies $f - 1 < \nu < f$, that carry edge currents. The incompressible (gapped) regions between the lines are at an exact filling $\nu = f$; they do not have low-energy charged excitations and so do not carry current, except when tunneling occurs. Tunneling through the energy barrier formed by the QHE gap occurs over a short distance $t$; for $t > 5\ell$ the tunneling rate $\propto \exp[-(t / 2\ell)^2]$ is exponentially small. Here, the magnetic length $\ell = \sqrt{\hbar / eB}$. Tunneling between different Landau levels does not conserve angular momentum, or involves a spin flip, and is expected to be much weaker; thus the backscattering via a "compressible island" ("type ii" in Ref. 17) is not considered here.

In 2D, a compressible QHE state is formed near half-filling, when the top Landau level is half-filled.[24] However, a confining potential lifts electron state degeneracy and a small confined "compressible island" is, in fact, incompressible in the low-temperature limit. This fundamental fact, and the following detailed considerations seem difficult to reconcile with forward scattering via a "compressible island" as the mechanism of the conductance oscillations reported in experiments. (i) Conductance oscillations periodic in applied uniform magnetic *field B* are observed in experiments. Gauge invariance requires periodicity in magnetic *flux* $\Phi = BS$;[3] a well-defined area $S$ is necessary to translate a uniform field into flux through this area. Aharonov-Bohm area is well-defined, but it is not clear what exactly is the area of a "compressible island". (ii) As a function of Landau level filling factor $\nu = nh/eB$, in a transition between QHE plateaus $f$ and $f+1$, the size of the "compressible island" changes from zero to a maximum value, so that a large variation of the $B$-period would result near half-filling $\nu = f + \frac{1}{2}$. (iii) Even at the maximum size, the radius of the "compressible island" must be less than the outer Aharonov-Bohm edge ring by at least $5\ell = 120$ nm at $B = 1.2$ T (cf. Fig. 2). Thus one would expect a 30% smaller "compressible ring" area and thus 30% different oscillation periods $\Delta_B$ if the two distinct mechanisms were involved. This is not seen in the experiment, the maximal variation of $\Delta_B$ is under 10% (see Fig. 2).

Similar conductance oscillations have been observed in devices with variously depleted constrictions, relative to the island center, from 5% to 50%. Different saddle point constriction depletion and filling factor would result in different edge channel structure in the island. In a device with 50% depletion, Landau level filling $\nu = 4.5$ in constrictions is accompanied by filling $\nu = 9$ at island center, so that several concentric compressible rings would be expected to form; while in a device with 5% depletion, the island center has $\nu = 1.26$ when $\nu = 1.20$ in constrictions, when oscillatory conductance has been reported, but no "compressible island" is expected at filling $\nu = 1.26$. Thus widely different regimes of constriction versus island center Landau level fillings $\nu$ result in similar oscillatory behavior.

While pleasingly simple and easy to visualize, the edge-channel models, like that of Fig. 4, have certain serious drawbacks. It can be deceptive to imply both tunneling rate and the QHE filling by one set of lines, while tunneling is exponentially sensitive to distance and thus to detail of constriction. For example, in Fig. 4, $\nu_C = 2 + \varepsilon$, hole forward-scattering in the inner $2 < \nu < 3$ edge channels is shown; but electron back-scattering between the outer $1 < \nu < 2$ channels, over a shorter distance in the perpendicular direction, is also easy to envision. The tunneling rate for forward-scattering can be extremely different for short and long constrictions, depending on device fabrication, while the back-scattering rate is about the same. Another drawback is that the "compressible island" in Fig. 4, $\nu_C = 2 + \varepsilon$, is not truly compressible: the electron state degeneracy is lifted by the confining potential. These energies can be estimated as the increment of the selfconsistent (screened) confining potential over the distance separating two consecutive island-circling basis orbitals, like in quantum antidots.[4,5] This energy is 60 mK in the interferometer of Ref. 20, in agreement with thermal excitation experiments. In the present device it is slightly lower, but still greater than temperature or excitation.

The continuous experimental oscillation sequence in Fig. 2 is consistent with a single physical mechanism, rather than a different mechanism for the different regimes in the edge-channel model of Fig. 4 back- ($\nu_C = 3 - \varepsilon$) and forward- ($\nu_C = 2 + \varepsilon$) scattering, and also at half-filling ($\nu_C = 2.5$). Such interpretation has been disputed in Refs. 21-23, where two physically different regimes, called "Aharonov-Bohm" for back-scattering and "Coulomb blockade" for forward-scattering have been proposed. The oscillatory behavior at half-filling has not been anticipated in Ref. 17. However, no qualitative discontinuity in the oscillation period or amplitude at half-filling is apparent in the data of Fig. 2, and single physics, the Aharonov-Bohm interference of interacting electrons in QHE regime, appears to fit all the regimes.

We acknowledge discussions with D. Averin, B. Halperin and B. Rosenow, and Wei Zhou for help in experiments. This work was supported in part by the National Science Foundation under grant DMR-0555238.